\documentclass[preprint,review,12pt]{elsarticle}
\usepackage{amssymb}
\journal{Physica C}
\begin{document}
\title{Suppression of electron correlations in the superconducting alloys of  Rh$_{17-x}$Ir$_x$S$_{15}$}
\author{H.R. Naren, Arumugam Tamizhavel, A.K. Nigam}
\author{S. Ramakrishan\corref{cor1}}
\ead{ramky@tifr.res.in, Tata Institute of Fundamental Research, Homi Bhabha Road, Mumbai-400005, India, Fax number:+912222804610}
\address{Department of Condensed Matter Physics and Materials Science, Tata Institute of Fundamental Research, Mumbai-400005, India}
\begin{abstract}
We have studied the effect of Iridium doping (Rh$_{17-x}$Ir$_{x}$S$_{15}$) in the Rhodium sites of the strongly correlated superconductor Rh$_{17}$S$_{15}$. Even at low levels of doping (x = 1 and 2) we see a drastic change in the superconducting properties as compared to those of the undoped system. We deduce that there is a reduction in the density of states at the Fermi level from reduced Pauli susceptibility and Sommerfeld coefficient in the doped samples. Moreover, the second magnetization peak in the isothermal magnetization scan (`fishtail') which was very prominent in the magnetization data of the undoped crystal is suppressed in the doped samples. The temperature dependence of  resistivity  of the doped crystals show a remarkably different behavior from that of the  undoped crystal with the appearance of a minima at lower temperatures, the position of which is fairly constant at different fields. Our data supports the notion that Iridium, which is a bigger atom than Rhodium expands the lattice thereby, reduces the electron correlations that existed due to the interaction between closer lying Rhodium atoms in the undoped system.
\vskip 12pt
\noindent
MS. Number~~~~~~~PACS Number: 71.20.Be,74.25.Bt, 74.25.Ha, 74.25.Ad\\
\end{abstract}
\begin {keyword}
electron correlations, Superconductivity, Rhodium sulphide
\end{keyword}
\maketitle
\section{Introduction}
Superconductivity in the mineral (Miassite) Rh$_{17}$S$_{15}$ (cubic, Pm3m) was shown to arise due to strong electron correlations (presumably due to the high density of Rh-4d states at the Fermi level) from magnetic, thermal and transport studies in a well characterized polycrystalline sample (with T$_c \approx$ 5.4~K) by some of us recently \cite{r1}. This is now confirmed by
the single crystal studies of Settai {\it et al} \cite{r2}. A vortex phase diagram in the same sample has also been established \cite{r3} via exploration of the anomalous variations in critical current density (J$_c$(H,T)) extracted from ac and dc magnetization measurements. The dc magnetization hysteresis loops show the presence of a very broad fishtail with unresolved structure commencing deep inside the mixed state and lasting upto the upper critical field H$_{c_2}$. Similar behavior is also observed in a high quality single crystal grown by us. The observation of T$^2$ dependence of resistivity, enhanced susceptibility, moderate Sommerfeld coefficient ($\gamma$), large value of the upper critical field and large heat capacity jump 
($\Delta$C /$\gamma$T$_C$$=$2)  suggest that Rh$_{17}$S$_{15}$ is a strongly correlated system. This conjecture is further substantiated by the estimated value of 2 for Wilson's coefficient ($ \pi^2~k_B^2~\chi(0)/3\mu_B^2~\gamma$) and 5~x~10$^{-5}$ for the Kadawaki-Woods ratio (A/$\gamma^2$). One way to achieve a moderate density of low-energy fermionic excitations (as seen by the  appreciable value of $\gamma$)  is from the large density of states of the narrow 4d band of Rh at the Fermi level \cite{r4}. This is supported by the structure since some of the Rh-Rh distances are smaller than those that exist in the pure Rh metal. In order to understand the strongly correlated superconductivity in Rh$_{17}$S$_{15}$, we have studied the effect of Ir substitution for Rh in this compound. 

\section{The Samples}
The samples were prepared by reacting Rh powder (99.9\% pure) and Ir powder (99.9\% pure) with Sulphur powder (99.999\% pure) by the same method as described in \cite{r1}. Powder X-ray diffraction confirmed the cubic structure. The lattice constant of pure Rh$_{17}$S$_{15}$ was found to be 0.99093(2) nm \cite{r1} while for Ir substituted samples, the lattice constants are estimated to be  0.99144(3) nm and 0.99183(3) nm from a Rietveld analysis. Clearly, Ir substitution leads to the expansion of the lattice of 
Rh$_{17}$S$_{15}$. The EPMA measurements indicate that the doped samples had a small distribution of dopant concentrations. On an average Ir$_{1}$Rh$_{16}$S$_{15}$ contains around 3 atomic percent of Ir and Ir$_{2}$ Rh$_{15}$S$_{15}$ around 6 atomic percent of Ir which are about the values we expect from our mixture stoichiometry. Since Rh exists in four different crystallographic sites in Rh$_{17}$S$_{15}$ \cite{r1} we cannot find out, from these preliminary measurements, which Rh atom is being substituted by Ir.

\section{Results and Discussion}
Fig.~\ref{rho} shows the temperature dependence of resistivity for pure Rh$_{17}$S$_{15}$ and those of Ir substituted samples measured in a home made dc resistivity setup at zero field.  Unlike the monotonically decreasing resistivity ($\rho$(T)) with the decrease in temperature in pure Rh$_{17}$S$_{15}$, the $\rho$(T) of Ir substituted samples shows a well resolved shallow minimum before the superconducting transition. The position of the minimum is at around 35 K for both the doped samples. There is no appreciable change in the temperature of the minima upto magnetic fields of 5 T. The origin of the  minima is not understood at the moment. However, it could be related to the weak localization effects which could arise due to the presence of disorder in the doped samples as evidenced by the increase in their resistivity values. In fact, the ratio of resistivities of doped to undoped increases from around 1.6 at room temperature to around 8 at 10~K.
Fig.~\ref{shc} shows a plot of C$_{p}$/T versus T$^{2}$ for the undoped and doped samples at zero field from 2K to 10K. These measurements were done on a commercial Physical Property Measurement System (PPMS, Quantum Design, USA). Fits to the equation C$_{p}$/T = $\gamma$ + $\beta$T$^{2}$ are also shown in the same figures where $\gamma$ is the electronic contribution and $\beta$ is the lattice contribution. Thus estimated $\gamma$ values in Ir$_{x}$Rh$_{17-x}$S$_{15}$ show a decrease with increase in x from around 105 mJ/mol-K$^{2}$ for x = 0 to around 80 mJ/mol-K$^{2}$ for x = 2 (actual values are tabulated later). This clearly indicates a substantial reduction of density of states at Fermi level and hence of electron correlations with increase in doping. From the $\beta$ values one can estimate the Debye temperature $\theta _D$ from 
$$\theta _D = (\frac{12\pi ^{4}N r k_{B}}{5\beta})^{1/3},$$
where N is the Avogadro Number, r is the number of atoms per formula unit and $k_{B}$ is the Boltzmann constant. The Debye temperature decreases from 424K to 327K as x increases from 0 to 2 indicating a softening of the lattice with doping.
The temperature dependence of the susceptibility ($\chi$(T)) in a field of 3 T, as measured by a commercial SQUID magnetometer (MPMS, Quantum Design, USA) of Ir doped samples is shown in Fig.~\ref{sus1} as contrasted with that of pure Rh$_{17}$S$_{15}$. Since none of Rh, S or Ir carry a magnetic moment we expect to see a temperature independent susceptibility (essentially Pauli susceptibility) in our samples. However, although the values of  $\chi$(T) are small, the data display a distinct temperature dependence similar to that seen in pure Rh$_{17}$S$_{15}$ \cite{r1}. There is a gradual increase in susceptibility as one goes to lower temperatures until the superconducting transition. The weak temperature dependence of $\chi$(T) could arise due to a sharp density of states at the Fermi level and we speculate that the mechanism is similar to that in V$_3$Si \cite{r5}. This conjecture is recently supported by the observation of temperature dependent Knight shift from the $^{103}$Rh-NMR experiment \cite{r6}. In fact, the values of susceptibility per Rhodium atom in pure Rh$_{17}$S$_{15}$ are of the same order as that of susceptibility per Vanadium atom in V$_3$Si \cite{r5} near the transition. The inset shows the variation of $\chi$(300 K) as a function of x in Rh$_{17-x}$Ir$_x$S$_{15}$. The susceptibility values at room temperature reduce by more than an order of magnitude from undoped sample to Ir$_{2}$Rh$_{15}$S$_{15}$ (actual values are tabulated later). This can be attributed to a reduced density of states at the Fermi level and supports our deduction from the C$_{p}$ data discussed earlier. \\ 

Fig.~\ref{fish} displays what is termed as the ``fishtail effect'' in the isothermal scan of magnetization as a function of field for  Rh$_{17-x}$Ir$_x$S$_{15}$ samples. It refers to a non-monotonic dependence of the width of the magnetic hysteresis loop on the magnitude of the external magnetic field. The fishtail effect (FE) or ''second magnetization peak anomaly'' found in magnetization loops is one of the still puzzling properties observed only in bulk samples, either homogeneous or granular, of single or polycrystals of high-T$_c$ as well as low-T$_c$ superconductors. Although this phenomenon has been studied intensively in recent years (see, for example, Refs. \cite{r7}-\cite{r11} and references cited therein),
there is still no general agreement about its origin. Mainly two approaches are used to
explain this phenomenon. In the first one, the fishtail effect is regarded as a manifestation
of a non-monotonic dependence of J$_c$ on B, where J$_c$ is the critical current density,
determined in neglect of relaxation processes, and B is the magnetic induction. This
approach has been used to discuss different mechanisms that could result in such a
dependence J$_c$ on B (see, for example, Refs. \cite{r7}-\cite{r11}). In another approach, magnetic relaxation, which causes the currents circulating in the sample to decrease below their
critical value, plays the main role in explaining the unusual form of the magnetic hysteresis
loop. In this case the fishtail effect is attributed to the non-monotonic H dependence
of the rate of magnetic relaxation. From Fig.~\ref{fish} one can see that the width of the fishtail  decreases as a function of x suggesting that the 
increased disorder due to Ir substitution leads to this suppression.

Finally, the temperature dependence of the upper critical field (H$_{c_2}$) is shown in the Fig.~\ref{hc2}. These H$_{c_2}$ values were estimated from susceptibility and magnetisation measurements. It is interesting to note that the value of H$_{c_2}$ decreases rapidly with the substitution of Ir for Rh in Rh$_{17-x}$Ir$_x$S$_{15}$. The reduction in the density of states at the Fermi level (as supported by other bulk measurements) seems to reduce the upper critical field values when Rh is substituted by Ir. 
\section{Conclusion}
Table 1 shows the values of various estimated parameters in the doped and undoped samples.
Earlier, \cite{r1}, we have conjectured that the electron correlations in Rh$_{17}$S$_{15}$ could be due to a high density of states of Rh-4d band at Fermi level arising due to strong Rh-Rh interactions. After Ir doping there is a lattice expansion as revealed by the X-ray data and hence an increase of Rh-Rh distances in the compound which could affect the interaction between Rh atoms. This seems to lead to a decrease of density of states at Fermi level as supported by the reduction in the susceptibility and the $\gamma$ values. The drastic decrease in H$_{c_2}$ could also be due to the same reason. Increase in Ir substitution also increases the level of disorder in the system and this seems to strongly increase the pinning in the system as revealed by the strong suppression of fishtail.

\newpage
\begin{figure}
\caption{\label{rho} Temperature dependence of resistivity of pure Rh$_{17}$S$_{15}$ and Ir substituted samples. The resistivity of pure Rh$_{17}$S$_{15}$ is 
displayed along the Y-axis and those of Ir substituted samples are shown in the Y1-axis. The inset shows the low temperature (10K$<$T$<$60~K) resistivity behavior.
The minima observed in Ir substituted samples are clearly seen in this inset.}
\end{figure}
\begin{figure}
\caption{\label{shc} A plot C$_P$/T vs T$^2$ from 2 to 10 K  of pure Rh$_{17}$S$_{15}$ and Ir substituted samples is shown. Here C$_P$ is the heat capacity in (mJ/mol-K) and T is the temperature. The solid straight lines are fits to an expression which is described in the text. The inset shows the dependence of the Sommerfeld coefficient ($\gamma$) as function of the concentration x in Rh$_{17-x}$Ir$_x$S$_{15}$ alloys.}
\end{figure}
\begin{figure}
\caption{\label{sus1}Temperature dependence of susceptibility of pure Rh$_{17}$S$_{15}$ and Ir substituted samples. The inset shows the variation of $\chi$(300 K)
as a function of x for Rh$_{17-x}$Ir$_x$S$_{15}$ alloys.}
\end{figure}
\begin{figure}
\caption{\label{fish} Isothermal scan of magnetization vs field of pure Rh$_{17}$S$_{15}$ and Ir substituted samples.}
\end{figure}
\begin{figure}
\caption{\label{hc2}Temperature dependence of upper critical  field (H$_{c_2}$) of pure Rh$_{17}$S$_{15}$ and Ir substituted samples.
H$_{c_2}$ values are estimated from isothermal magnetization measurements.}
\end{figure}
\newpage
\begin{table}
\caption{Structural and thermal properties of Rh$_{17-x}$Ir$_x$S$_{15}$ system.}
\label{table 1}
\end{table}
\end{document}